%% file: revised.tex
\documentstyle[aps,bezier,amsmath,epsfig]{revtex}
\begin{document}

\draft

\title{Two-time scales, two-temperature scenario for nonlinear rheology}
\author{Ludovic BERTHIER$^{\ddag,\star}$, Jean-Louis BARRAT$^\star$
and Jorge KURCHAN$^\ddag$.}

\address{$^\ddag$Laboratoire de Physique \\ ENS-Lyon and CNRS,
 F-69364, Lyon Cedex 07, France }

\address{$^\star$D\'epartement de Physique des Mat\'eriaux \\
 Universit\'e Claude 
Bernard and CNRS, F-69622 Villeurbanne Cedex, France}

\date{\today}

\maketitle

\begin{abstract}
We investigate  a general scenario for 
``glassy'' or ``jammed'' systems driven by an external,
non-conservative force, analogous to a
shear force in a fluid. In this scenario,
the drive results in the suppression
of the usual aging process, and the correlation and response functions
become time translation invariant. The relaxation time
and the response functions are then dependent 
on the intensity of the drive and  on temperature.
We investigate this dependence 
within the framework of a dynamical closure approximation
that becomes exact for disordered, fully-connected  models.
The relaxation time is shown to be a decreasing function of
the drive (``shear thinning'' effect).  The correlation functions
below the glass transition temperature ($T_c$) display a two-time scales  
relaxation pattern, similar to that observed at equilibrium 
slightly above $T_c$. 
We also study the violation of the fluctuation dissipation
relationship in the driven system. This violation is very reminiscent of the
one that takes place in a  system {\em aging} below $T_c$ at zero drive.
It involves in particular the appearance of a 
 {\em two-temperature} regime, in the sense of an effective 
fluctuation dissipation temperature \cite{leticia3}.
Although our results are in principle limited to the 
closure relations that hold for mean-field models,
 we  argue that  a number of the salient
features are not  inherent to the approximation scheme,
 and may be tested in experiments and simulations.

\end{abstract}

\pacs{LPENSL-TH-18/99}


\section{Introduction}
The behavior of complex systems subject to an external
drive is a very complex field of research.
The generic situation we may think of is the case of 
a fluid undergoing steady shear flow, in which case
the study of its response pertains to the field of rheology. 
Rheological experiments on complex  fluids
are known to display a rich  phenomenology, as illustrated in 
the recent book by Larson \cite{larson}. 
Non-trivial behaviors, however,  are not restricted to complex fluids, since 
it is also found that supercooled liquids exhibit a non-Newtonian
``shear thinning'' behavior \cite{onuki}.

From a fundamental and theoretical point of view, the 
most interesting features emerge when the intrinsic relaxation
time scale of the system become of macroscopic order, 
so that there is a direct interplay between the  ``shearing''
time scale and the relaxation time scale. 
This means that we will be interested
in systems whose  dynamical evolution exhibits a {\em two-time scales}
pattern. This is the case e.g.  in  supercooled liquids, in
which the particles have a fast ``rattling'' motion  inside the ``cage'' 
constituted by their  neighbors, followed by  a slow ``structural'' 
 rearrangement of these cages. The two time scales
become more and more different on approaching  the glass transition.
Below the glass transition the same behavior subsists, but now the
time scale of the structural relaxation (the $\alpha$-relaxation time
 $t_\alpha$)
is not constant and grows
with the waiting
time elapsed after the quench to low temperatures: 
this gradual arrest is called aging. 
More generally, 
 the same situation
is realized in all systems that are ``jammed'', like  granular
materials or foams \cite{jorge}.
In all these cases, it is
 known that a driving force has
a particularly strong influence, for it may be able to stop the aging
in an out of equilibrium system and to restore time translation invariance.

In order to study theoretically this interplay between the drive and
the relaxation of the system, it is natural to extend the
approaches that have been successful for describing the 
dynamical behavior of glassy systems, 
as reviewed e.g. in \cite{review_aging}. 
One possible and quite promising  route, followed by Sollich 
and coworkers (for a review, see \cite{sollich} and references therein)
is to extend the phenomenological ``trap'' models to driven systems,
giving rise to the so-called SGR model. Such a model was shown to 
account well for a number of the generic features found in soft
glassy materials.

Another possible, and complementary, approach
is to extend the mode-coupling approach 
used in the study of glassy systems to the driven case. 
This approach relies on a closure relation for the dynamical 
equations that is known to be exact for ``mean-field'' like systems, 
and to provide a reasonably good description
of the dynamics of real glass forming liquids \cite{gotze}. 
This theoretical framework   has the 
advantage of simultaneously giving insights 
into   the macroscopic (rheological)
and the microscopic (non-equilibrium statistical mechanics) aspects of
the problem, since the mode-coupling equations can be derived from
specific microscopic models.

Such microscopic models with driving forces 
have  been first studied in the context of neural networks
\cite{Grheso,CS}, the drive being a tool to destroy the glassy phase.
More closely related to our approach  are the studies of Horner \cite{horner}
and Thalmann \cite{fabrice} who investigated 
the dynamics of a particle in a random potential in the 
presence of a driving force. Our focus in this paper
will be a bit different, since we are interested in translating
the results into the language of non-linear rheology.

The paper is organized as follows. In the next section we 
explain the general spirit of the closure approximations that give rise
to the mode-coupling equations. An explicit example 
for a simple system is worked out, and the numerical methods
used in solving the equations are presented. 
Section III contains the 
results obtained for the correlation and response functions 
in a stationary driven state. We discuss our results in section IV,
and conclude by considering  some
possible extensions of this work in section V.

\section{Dynamical equations}

\subsection{The general framework}

 We   start by    briefly describing \cite{bouchaud} the 
perturbative resummation schemes (of which the mode-coupling approximation
is a special example) that allow to obtain closed equations for the
time dependent correlation and response functions of an interacting system. 
We  then specialize  to the simpler case of a single mode, 
and study the resulting equations numerically.

Let us consider a system whose dynamics is described at the microscopic level
by  a Langevin equation
\begin{equation} m \frac{\partial^2
{\bbox \phi}({\bbox x},t) }{ \partial t^2} 
+
\frac {\partial
{\bbox \phi}({\bbox x},t) }{ \partial t} 
= 
(-\mu(t)+\Delta) {\bbox \phi}({\bbox x},t) - g {\bbox F}({\bbox \phi})
 + {\bbox \eta} ({\bbox x},t)
+ {\bbox h}({\bbox x},t) ,
\label{langevin}
\end{equation}
where ${\bbox \phi}({\bbox x},t)$ is a vector field, ${\bbox F}({\bbox
\phi})$ 
is a non-linear 
(possibly  non-local) 
coupling term and ${\bbox \eta}$ a Gaussian white noise.   The term
containing $\mu(t)$ 
is a restoring force and $\Delta$ may contain spatial 
  derivatives or  convolutions 
(as one would obtain for instance if $\phi$ is a coarse-grained density and 
the evolution is driven by the Ramakrishnan-Youssouf density functional,
see e.g. \cite{dasgupta}).
The coupling constant $g$ 
serves as a book-keeping parameter 
to set up a perturbative expansion.

Making a spatial Fourier transform, we obtain the fields 
 $ {\hat \phi}({\bbox k},t)$
and the Gaussian noise such that $\langle \eta({\bbox k},t) \eta({\bbox
k'},t')\rangle = 2 T \delta({\bbox k+\bbox k' })
  \delta(t-t')$. 
In terms of those Fourier transformed variables, the  correlation
 and response functions are defined as
\begin{eqnarray}
\delta({\bbox k}+{\bbox k'}) C({\bbox k},t,t') 
&=& \langle \hat \phi({\bbox k},t) \hat \phi({\bbox k'},t') \rangle ,
\\
\delta({\bbox k}+{\bbox k'}) R({\bbox k},t,t') &=& 
\langle \frac{\delta \hat \phi({\bbox k},t) }{\delta{\bbox h}
 ({\bbox k'},t')} \rangle .
\end{eqnarray}
The dynamical equation (\ref{langevin}) implies a Dyson equation
for $C$ and $R$ of the form
\begin{eqnarray}
m \frac{\partial^2 C({\bbox k},t,t')  }{\partial   t^2}+
\frac{\partial C({\bbox k},t,t')   }{\partial t}  &=& (- \mu
(t)\delta({\bbox{k'}})+\Delta_{\bbox k'}) C(\bbox{k-k'},t,t')
+ 2 T R({\bbox k},t',t) 
\nonumber \\
& &+ \int_{
-\infty}^{t'} dt'' D({\bbox k'},t,t'')R(\bbox{k-k'},t',t'') +
\int_{-\infty}^{t} dt'' \Sigma({\bbox k'},t,t'') C({\bbox{k-k'}},t'',t') ,\\
m \frac{\partial^2 R({\bbox k},t,t')}{\partial t^2}+
\frac{\partial R({\bbox k},t,t')}{\partial t}  
&=& (-\mu (t)\delta({\bbox{k'}})+ \Delta_{\bbox k'}) R({\bbox{k-k'}},t,t')
 + \delta(t-t') \nonumber \\
& &+ \int_{
t'}^{t} dt'' \Sigma({\bbox k'},t,t'') R({\bbox{k-k'}},t'',t') 
\label{dyson}
\end{eqnarray}
where the summation over $\bbox{k'}$ is implied.
The operator $\Delta_{\bbox k}$ acts now in the Fourier space.
The functions  $\Sigma({\bbox k},t,t')$ and $D({\bbox k},t,t')$ can be
obtained \cite{cyrano} as {\em functionals} of 
the correlations and responses 
by adding all the  
two-line irreducible diagrams of the perturbative  expansion in $g$
of (\ref{langevin}) but substituting the propagators 
with the dressed propagators   
$C({\bbox k},t,t')$ and $R({\bbox k},t,t')$.
Now, if we stay at the level of the simplest diagrams (having only two
vertices), we obtain the simplification\footnote{This is a resummation
  and is not the same
  as expanding only to order $g^2$, since the propagator themselves depend
  also on $g$.} that 
 $\Sigma({\bbox k},t,t')$ and $D({\bbox k},t,t')$
become ordinary {\em functions}  of 
$C({\bbox k},t,t')$ and $R({\bbox k},t,t')$ (with no integrations over
the times). This type of 
ansatz constitutes the basis of the mode-coupling approximation.

Naively, one could expect
that the mode-coupling strategy could be improved
to any desired accuracy
by including  diagrams  with higher order vertices.
Unfortunately, it seems that the phenomena
usually described as ``activated processes'' 
are of a non-perturbative nature, and hence will be 
missed even if higher order vertices  are taken into account.
This is an intrinsic limitation of the mode-coupling or mean-field
 approaches when applied to realistic systems, 
and has to be taken into account when interpreting
the results.
We shall discuss this point in a more detailed way at the end
of Section II D.

\subsection{A single-mode driven model}
\label{approx}

Staying within this approximation, we furthermore
make the considerable simplification of considering a single $k$-mode.
This of course means that we  give up all spatial information.
The basic elements we  find, however, can be readily generalized 
to the case of many   $k$-modes, and, furthermore, to
approximations that include  diagrams with more and more vertices.
The spirit is similar to the study of ``schematic models'' for the
theory of supercooled liquids by G\"otze and coworkers \cite{gotze}.
These models are also closely related to spin glass ones, as explained
in the next section. 
No attempt is made 
to describe in a  {\it quantitative} way the rheology of 
glassy systems, but still it is hoped
that {\it generic} and non-trivial behaviors 
can be predicted at a qualitative level.

If we restrict equation (\ref{dyson}) to a single ``important''
mode, and  furthermore absorb $\Delta$ which is now irrelevant in $\mu$,
and neglect 
the inertial term which is inessential for the slow dynamics, we
obtain
\begin{equation}
\begin{aligned}
\frac{\partial C(t,t')}{\partial t}  = & - \mu (t) C(t,t') + 2 T
R(t',t) + 
\int_{-\infty}^{t'} dt'' D(C(t,t''))R(t',t'') + 
\int_{-\infty}^{t} dt'' \Sigma(t,t'') C(t'',t') 
,\\
\frac{\partial R(t,t')}{\partial t}  = & 
-\mu (t) R(t,t') + \delta(t-t') + \int_{t'}
^{t} dt'' \Sigma (t,t'') R(t'',t') .
\label{eqs}
\end{aligned} 
\end{equation}
If in addition, we impose $C(t,t)=1$  we have
\begin{equation}
\mu(t)  =  T + \int_{-\infty}^{t} dt'' 
\left[  D(C(t,t'')) R(t,t'') + \Sigma(t,t'') C(t,t'') \right].
\end{equation}

When the force in the Langevin equations 
derives from a potential, {\em so that detailed balance is verified},
one has \cite{bouchaud}
\begin{equation}
\Sigma(t,t')=D'(C(t,t'))R(t,t'),
\label{balance}
\end{equation}
with $D'(x)\equiv dD(x)/dx$. 
Conversely,  a  set of equations  with  
$\Sigma(t,t')-D'(C(t,t'))R(t,t') \neq 0$ can only
describe a driven system, in which detailed balance is violated.
Mode-coupling equations that describe a driven system can therefore 
be obtained  by introducing a modified version of
equation (\ref{balance}).
Cugliandolo {\it et al} \cite{leticia2} chose for instance
\begin{equation}
\Sigma(t,t')=\alpha D'(C(t,t'))R(t,t'),
\end{equation}
the parameter $1-\alpha$ being then a measure of 
the (non-conservative)
driving forces.
They showed
numerically that the presence of the drive was sufficient to stop 
aging in the glassy phase, so that time translation invariance was recovered
at all temperatures.

It may be important at this point to distinguish between two ways 
of driving a system:
 
({\em i}) ``Shear-like'' driving: the system is subjected to forces that
do not
derive from a global potential, as when a potential difference is
applied at the ends of a conductor and the circuit is closed.
 They can be time-dependent or constant,  
and generate currents in both cases.

({\em ii}) ``Tapping-like'' driving: the forces are time-dependent but do
derive from a global potential. This is for example the case of an
a.c. magnetic field in spin models,
 or oscillating acceleration  of a container with
frictionless walls (an a.c. gravity
field).
These forces do not do work if they are constant in time,
independently of their strength.

In this paper we concentrate on the case of continuous drive, 
and hence only the ``shearing-like'' forces are relevant. We discuss
below how the analogy with a rheological experiment can
be developed further. The case of
an oscillating drive will be discussed elsewhere.

\subsection{The associated disordered model, and the rheological analogy.}
\label{pspin}

Many years ago, Kraichnan \cite{kraichnan} noted that one could find 
a disordered model such that the approximate closed 
equations for the two-point correlations and responses of the original
model are  {\em exact} for it. Although this hidden model behind the closure
approximation is not otherwise directly related to the original one,
it allows to view the dynamical equations from a different,
instructive perspective.

We now specify the model we concentrate on in the rest of the paper.
The case of the single mode equations with $D(x)=px^{p-1}/2$
corresponds to a 
disordered model given by continuous variables $s_i$ 
($i=1,\cdots,N$) evolving with the Langevin equation
\begin{equation}
\frac{\partial s_i (t)}{\partial t} = -\mu(t)s_i(t) -
 \frac {\delta H}
{\delta s_i(t)} + f_i^{\text{drive}}(t) 
+ \eta_i(t),
\label{single}
\end{equation}
where
\begin{equation}
H = - \sum_{j_1 < \cdots < j_p} J_{j_1 \cdots j_p} s_{j_1}
 \cdots s_{j_p} 
\label{ener}
\end{equation}
is the Hamiltonian of the $p$-spin model and 
\begin{equation}
 f_i^{\text{drive}} =  \epsilon(t) \sum_{i}^{\ast} 
\tilde{J}_i^{j_1 \cdots j_{k-1}} s_{j_1} \cdots s_{j_{k-1}},
\label{force}
\end{equation}
with $ \sum_{i}^{\ast} \equiv \sum_{i<j_1 \cdots < j_{p-1}} + 
\sum_{j_1< i < j_2 \cdots < j_{p-1}} + \cdots + \sum_{j_1< 
\cdots < j_{p-1} <i}$.
The parameter $\mu(t)$ ensures a spherical constraint 
$ \sum_i s_i^2 =N$, and $\eta_i(t)$ ($i=1,\cdots,N$) 
are random Gaussian variables with mean 0 and variance $2T$. 
The couplings  $J$ in $H$ are random Gaussian variables, symmetrical
about the permutations of $(j_1,\cdots,j_p)$, 
with mean zero and variance 
$p!/2 N^{p-1}$.
The couplings $\tilde{J}$ in $f_i^{\text{drive}}$ are random Gaussian
variables, symmetrical about the permutations 
of $(j_1,\cdots,j_{k-1})$, with mean zero
and such that
\begin{equation}
\overline{\tilde{J}_i^{j_1 \cdots j_{k-1}} \tilde{J}_i^{ 
j_{1} \cdots j_{k-1}}} = \frac{k!}{2N^{k-1}} ; \quad
\overline{ \tilde{J}_{i}^
{j_1 \cdots j_{k-1}} \tilde{J}_{j_r}^{ j_1 
\cdots  i \cdots j_{k-1}}} =  0.
\end{equation} 
The resulting force cannot be written as the
derivative of a potential.

Equations (\ref{eqs}) are then the exact equations satisfied by 
$C(t,t')= \sum_i <s_i(t) s_i(t')>/N$ \\ and  
$R(t,t')= \sum_i < \delta s_i(t) / \delta \eta_i(t')>/N$ 
in the limit $N \rightarrow
\infty$ and with $C(t,t)$ imposed.
The model studied in Ref. \cite{leticia2} corresponds to $k=p$.

The important point is that, 
on average, {\it only the non-conservative part of 
the force gives energy to the system}, hence the name ``driving force''.
The  amplitude of the drive is 
controlled by the parameter $\epsilon(t)$. If we now want to push further
the analogy with the dynamics of a fluid system undergoing shear flow,
we have to identify the equivalent of the shear rate and stress variables. 
Obviously, 
the schematic character of the model makes it difficult 
to carry out such an 
identification at a microscopic level. A way of bypassing this 
difficulty is to 
estimate the power input into the system due to the existence of 
the non-conservative
forces, which can be defined as
\begin{equation}
P \equiv \overline{\langle \frac{1}{N} \sum_{i=1}^N   f_i^{\text{drive}}
\dot{s_i} \rangle }.
\label{puissance.def}
\end{equation}
The calculation of this quantity is given in detail in the appendix, 
and assuming stationarity (see below) yields
\begin{equation}
P= \frac{k \epsilon^2}{2} \int_0^{+\infty} d\tau C(\tau)^{k-1} 
\frac{d R(\tau)}{d \tau}.
\label{eq15}
\end{equation}
If we assume that the fluctuation dissipation theorem is not
 violated too strongly, as will be the case in the following
 examples, $R$ is roughly proportional to $dC/dt$.
Hence the power dissipated in the system will scale as
 $\epsilon^2/t_\alpha$, 
where $t_\alpha$ is the relaxation time of the system. 
Comparing to a standard shear flow, this 
indicates that $\epsilon$ should in that case be interpreted 
as playing the role of a stress,
while $\epsilon/ t_\alpha$ is analogous to a shear rate.

Although equation (\ref{eq15}) can be used to define the analogous
of shear rate and shear stress in our model, the response
function $R$ does not have a direct analog in rheological measurements,
since it is not the response associated to a shear stress.
Experimentally, one should think of $R$ and $C$ as the response and 
correlation functions associated to an observable which is not
rheologically relevant (as measured e.g. in a dielectric measurement).
This observable would be measured in a system made stationary
by imposing an external shear rate.

\subsection{Reynolds effect and yield stress}
\label{yield}

 Having access to a microscopic model behind the  dynamical equations 
allows us to understand the evolution  from 
the geometry of  the corresponding phase-space. 
At zero external drive, this connection has been studied  in much 
detail  \cite{Kupavi,Crso,Anire,ABarrat}, 
and the main results can be summarized as follows.

At a given temperature,  the free energy
 landscape of the purely conservative model with energy (\ref{ener})
can be constructed. Above a dynamical transition
temperature $T_c$,
the available phase space is dominated by one large
basin in the free energy, corresponding to the paramagnetic 
(or ``liquid'') state. At $T_c$, a {\em threshold level} in free
 energy appears, below
 which the  free-energy surface is split into  
 exponentially many disconnected regions. 

The aging dynamics below $T_c$ can be understood \cite{Cuku} as a
gradual descent  to the threshold level, starting from  high energy
configurations.
  The slowing down is then  the
consequence of the decreasing  connectivity of the visited landscape.
 The system never really reaches the deeper, very
disconnected parts of phase-space below the threshold.
On the other hand, if the system is somehow prepared in one of these
deep regions, it remains trapped there for all times.

When the system is quenched from a high temperature, but at the same
time driven by non-conservative forces, it  remains drifting  above
the free energy threshold, constantly receiving energy from the drive. 
In a mean-field system, such a situation will hold for  arbitrarily weak
drive, since the undriven system itself never falls below
this energy level.

If, on the contrary  the system is prepared in an energy 
state below the threshold,
we expect that  a weak
driving force will have essentially no effect (beyond an ``elastic''
response of the system),
as it is not strong enough to make the
system overcome the barriers. If instead a strong drive is applied, the
system escapes the low-lying valley and it surfaces above the
threshold,
where the  drive will suffice to keep it forever
(recall that in the rheological analogy, by ``drive'', we really
mean a stress).
An interesting 
analogy can be drawn with the Reynolds dilatancy 
 effect \cite{jorge}
in granular materials, or with the yield stress in a 
Bingham fluid \cite{larson}. Note that
the actual
value of valley depths and threshold level can be calculated 
from first principles within the same scheme of approximation
we have used for the dynamics, starting from the original microscopic model.

We are now in a position to see what is the main shortcoming
 of the mode-coupling
kind of approach. In any  realistic system the structure of threshold and
valleys may remain essentially the same, but now activated
(non-perturbative)
processes allow to jump barriers that are impenetrable at the
perturbative level.
Hence, for a real  system the relaxation (aging) process 
will allow the system to  penetrate slowly below the threshold, and 
after long times to access  deeper valley. In order to keep
the system with activated processes from sinking, we now need driving
forces with  finite strength (the ``yield stress'' effect).
This effect can be expected to take place whenever the system
has fallen out of equilibrium, and is undergoing an aging process.
Aging dynamics leads the system to be trapped (on the experimental time scale)
in some deep lying 
valley. A strong enough drive can 
 force  to leave the valley, and 
 a small extra drive is sufficient to  
keep the system above the threshold.

\section{Results for a steady drive}

\begin{figure}
\begin{center}
\input{corrT=.613.tex}
\vspace*{0.5cm}
\caption{Correlation function vs. time at $T=0.613>T_c$ for different
driving forces. From left to right: $\epsilon = 5$, 0.333, 0.143, 
0.05, 0.0158, 0.00447 and 0.
The longest plateau corresponds then to the undriven case.}
\label{corrT=.613}
\end{center}
\end{figure}
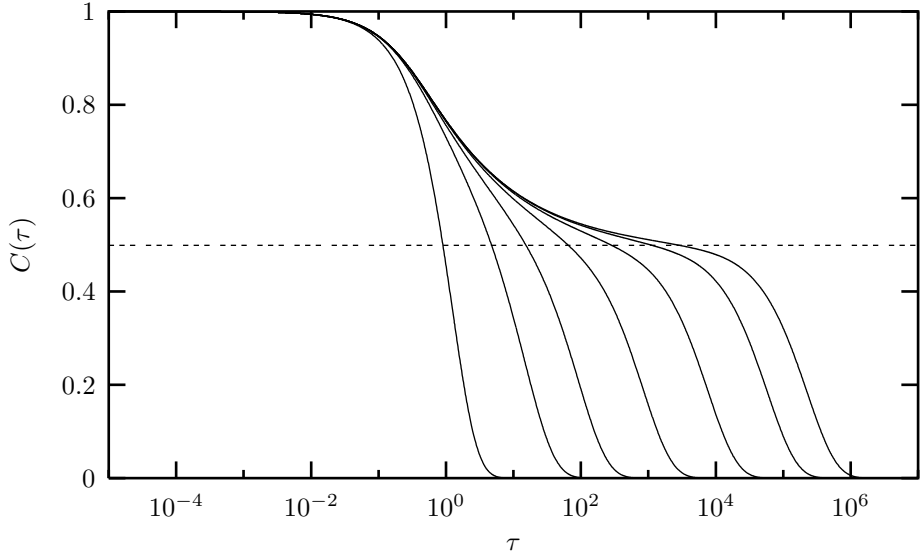 

\subsection{Dynamical equations in the stationary state}

When the system is submitted to a steady drive, it eventually reaches
a stationary state, whatever the temperature.
This allows to replace in equations (\ref{eqs}) the two-times functions
$A(t,t')$ by $A(\tau)$ with $\tau \equiv t-t'$.
The following equations corresponding to the simple model
described in  section \ref{pspin} are easily obtained: 
\begin{equation}
\begin{aligned}
\frac{d C(\tau)}{d \tau}  = & -\mu  C(\tau) + \int_{0}^{\tau} d\tau' 
\Sigma (\tau -\tau')  C(\tau')  + \int_{0}^{+\infty} d\tau' 
\left[ \Sigma (\tau+\tau') C(\tau')  + D(\tau + \tau') 
R (\tau') \right], \\
\frac{d R(\tau)}{d \tau}  = & -\mu  R (\tau) + \int_{0}^{\tau} d\tau' 
\Sigma (\tau-\tau') R(\tau') ,\\
\mu  = &  T +  \int_{0}^{+\infty} d\tau' \left[  D(\tau') 
R (\tau') + \Sigma (\tau') C(\tau') \right] ,\\
D(\tau)  = & \frac{p}{2} C(\tau)^{p-1} + 
\epsilon^2 \frac{k}{2} C(\tau)^{k-1}, \\
\Sigma (\tau) = & \frac{p(p-1)}{2} C(\tau)^{p-2} R(\tau).
\label{system}
\end{aligned}
\end{equation}
Note that these equations are ``non-causal'' in the time difference
$\tau$.
They are of course still causal in the original two times,
but the parity of $C(\tau)$ has now been used.

The system (\ref{system}) can only be solved analytically
in the limit of small drive, for all temperatures,
following the steps of Ref. \cite{Cuku}.
The solution shows that 
the correlation and response functions can be split 
 into two parts associated
with the two time scales discussed in the introduction.
One then writes 
$C(\tau)=C_s(\tau)+C_f(\tau)$ and $R(\tau)=R_s(\tau)+R_f(\tau)$,
`$f$' (`$s$') labeling the `fast' (`slow') time-scale.
In that small drive limit, the two time scales  are well separated and the 
fluctuation-dissipation theorem (FDT) does not hold.
More precisely, one can prove  a generalization of the FDT, in the form
\begin{equation}
R_f(\tau) = - \frac{1}{T} \frac{dC_f(\tau)}{d\tau};\quad
R_s(\tau) = - \frac{1}{T_{\text{eff}}} \frac{dC_s(\tau)}{d\tau}.
\end{equation}
An interpretation is that the short time scale is thermalized at the 
bath temperature $T$, while
the longer one is thermalized at an effective temperature
 $T_{\text{eff}}$ \cite{leticia3}.
This effective temperature is determined analytically by the matching
of the solutions between the two time scales.
It is interesting to remark that  the values of the effective
temperatures for a stationary, very weakly driven system coincides
with the effective temperature of its undriven, aging counterpart.

In order to study the pre-asymptotic ($\epsilon \neq 0$) behavior 
of the system, we solved numerically the system (\ref{system}), combining the 
numerical methods of Refs. \cite{ng,arnulf}.
In the next section, we discuss  the numerical results obtained for the case 
$k=p=3$.  The cases $k=2$, $p=3$ and $k=p=4$ have also been studied
 numerically,
with very similar results. We discuss in two different subsections the results
for $T>T_c$ and $T<T_c$.

\subsection{Above the glass transition temperature, $T>T_c$ }

\begin{figure}
\begin{tabular}{cc} 
\psfig{figure=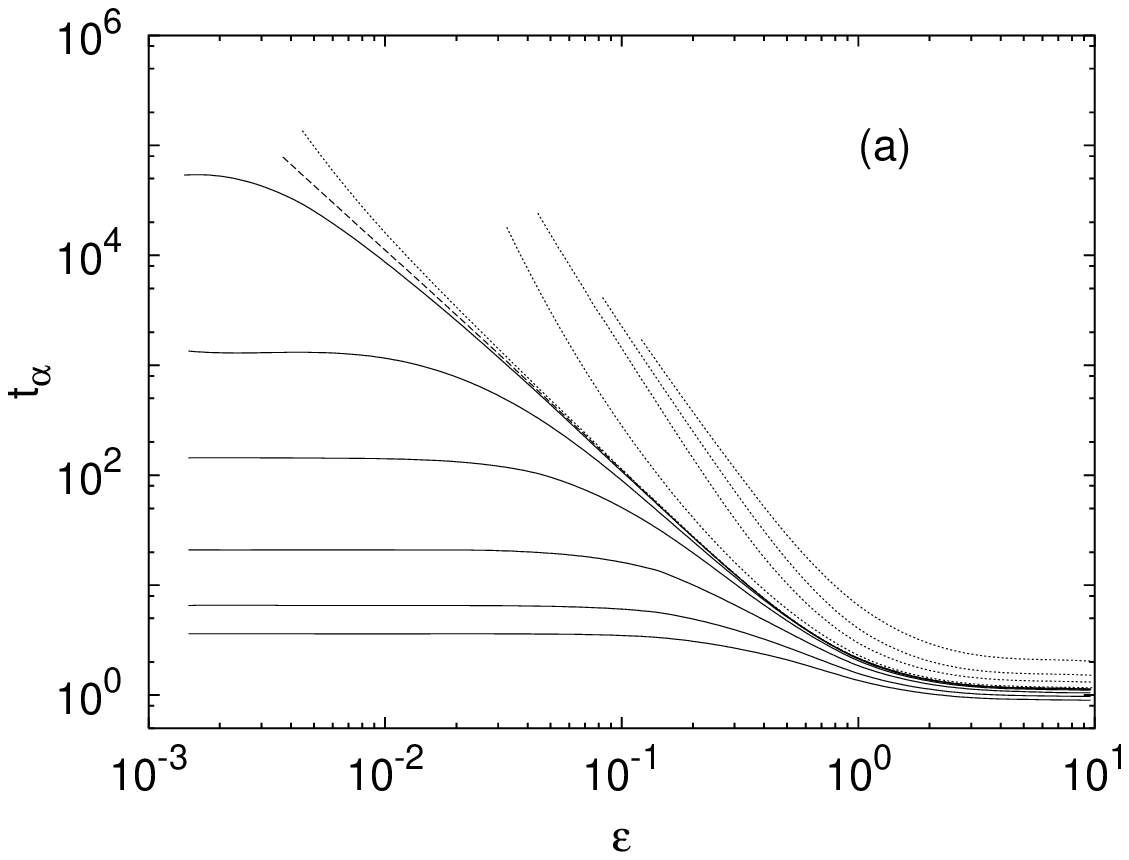,height=7.5cm,width=8.5cm} & 
\hspace*{-1cm}
\psfig{figure=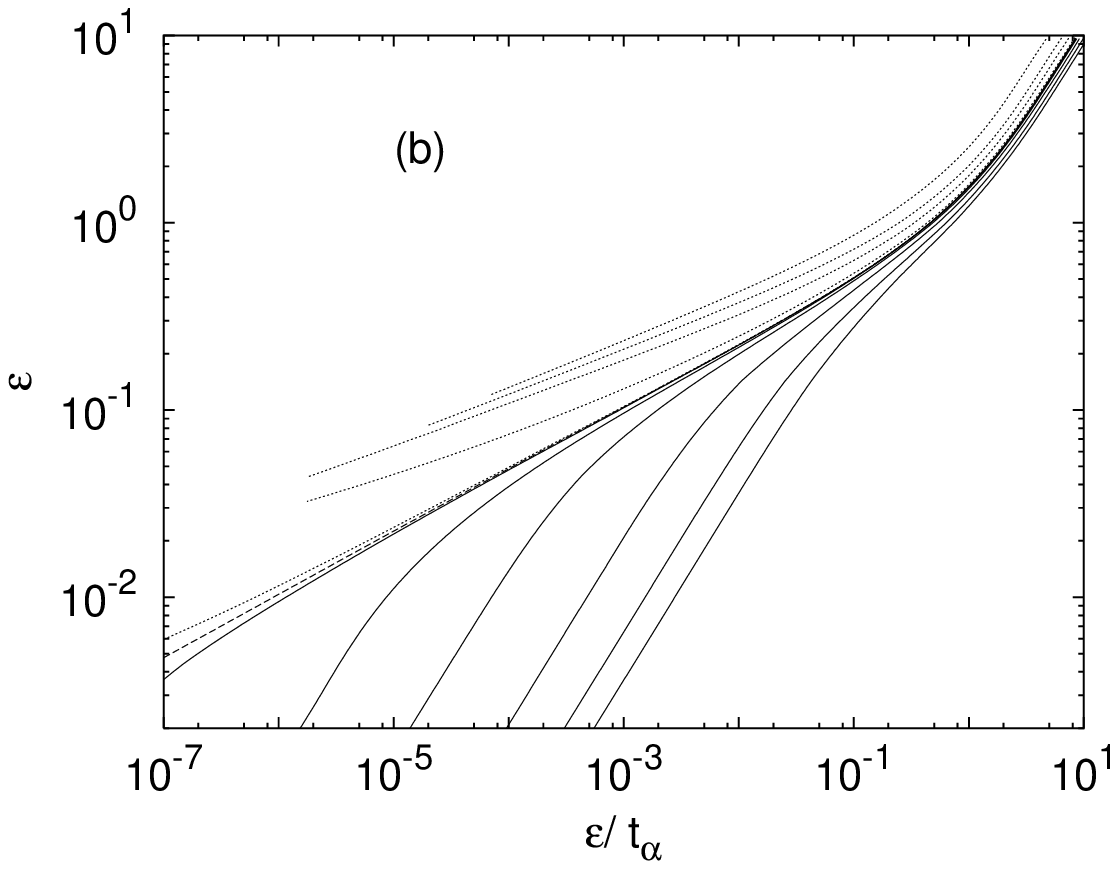,height=7.5cm,width=8.5cm}
\end{tabular}
\caption{(a) Alpha-relaxation time as a function of drive
for temperatures (from bottom to top) 
 $T=0.9$, 0.8, 0.7,
0.64, 0.62, 0.613, $T_c \simeq 0.61237$, 0.6115, 0.58, 
0.45, 0.3, 0.01.
Full lines are for temperatures above $T_c$, the dashed line is $T=T_c$,
and the dotted lines are for $T<T_c$.
(b) Flow curves analogous to the usual $\sigma$ vs $\dot{\gamma}$ 
for the same temperatures in the same order.}
\label{div}
\end{figure} 

The correlation function decay  above the glass transition temperature is a
two-step process, as shown in figure \ref{corrT=.613}. 
The length of
the plateau (the $\alpha$-relaxation time $t_\alpha$)
increases as the glass transition temperature is approached. 
At fixed temperature, for stronger driving forces
the plateau region becomes shorter, and for a
sufficiently
strong drive the process has  no longer two distinct time scales. 
In what follows, we  focus on the  not too strong drive
regime in which two steps are still discernible.

The full lines in figure \ref{div}-a show
the variations of $t_\alpha$  as a function of the 
amplitude of the driving force $\epsilon$ above $T_c$. 
The relaxation time 
is defined here as  $t_\alpha (\epsilon,T) \equiv
 \int_0^{+\infty} d\tau C(\tau)$. In the rheological analogy,
$t_\alpha$  is expected to play the role of a viscosity 
(which, generally speaking,
scales as the structural relaxation time).
Hence these curves are somewhat analogous to the 
``flow curves'' measured in non-Newtonian fluids,
and they give the behavior of the viscosity as a function
of the shear stress.
The translation to the more common plot $\sigma$ (for us $\epsilon$)
as a function
of $\dot{\gamma}$ (for us $\epsilon/t_\alpha$) is done in figure 
\ref{div}-b.

Above the glass transition ($T>T_c$),
$t_\alpha$ levels off  at a finite value $t_0(T)$ when the drive vanishes,
except at exactly $T_c$ where it diverges like $\epsilon^{-\beta}$
(the value of $\beta$ is discussed below).
The Mode-Coupling Theory \cite{gotze} gives the scaling of $t_0$ when 
approaching $T_c$ and, for the model studied here, implies
$t_0 (T) \sim (T-T_c)^{-1.765}$, which is very well verified numerically.
Above $T_c$, the curves of figure \ref{div}-a 
exhibit a plateau for small driving forces whose
height diverges as $T \rightarrow T_c$. 
This plateau corresponds to a Newtonian
regime in the language of rheology, 
sometimes also called ``linear regime''.
By this we mean that the relaxation time 
in this region (small values of $\epsilon$) is essentially independent of the 
drive, implying that the drive does not substantially alter the 
dynamics of the system.

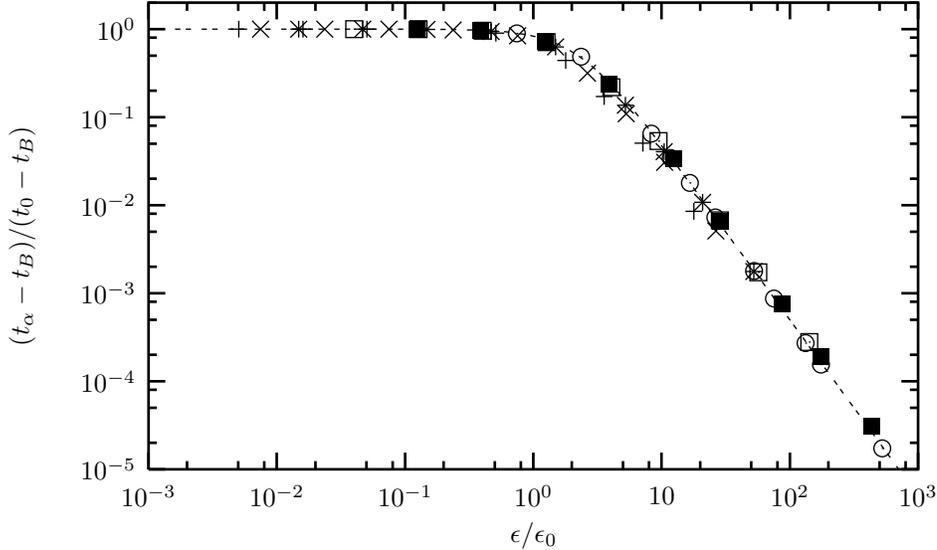
\begin{figure}
\begin{center}
\input{scaleaboveTc.tex}
\vspace*{0.5cm}
\caption{Scaling plot of $t_{\alpha}$ versus drive for several
temperatures above the transition $T=0.9$, 0.8, 0.7, 0.64, 0.62 and 0.613.
The dashed line is the fit $1/(1+x^2)$.}
\label{scaleaboveTc}
\end{center}
\end{figure}

For stronger driving forces, the relaxation time decreases,
a {\it shear thinning} effect.
The expression 
\begin{equation}
t_\alpha (\epsilon,T) = t_B + \frac{t_0 - t_B}
{1+(\frac{\epsilon}{\epsilon_0})^\beta}
\label{fit}
\end{equation}
fits the relation $t_\alpha(\epsilon,T>T_c)$ very well
as can be seen in  figure \ref{scaleaboveTc}.
The  fit is obtained (following Ref.\cite{ferry}), by  defining
$t_B(T) \equiv \lim_{\epsilon \rightarrow \infty} t_\alpha$, and
the second fitting parameter $\epsilon_0$ as 
$t_\alpha (\epsilon_0) \equiv 0.8 t_0$. Numerically,
we find  $\beta = 2$. Interestingly, the same $\beta$
is found for the other cases (corresponding to $k=2, p=3$ and $k=p=4$)
we have considered. 
We note that such an exponent is clearly non-trivial,
in the sense that it could not be predicted from a simple 
dimensional argument.

The shear thinning effect is well documented 
 in systems as different as gels, polymers or
supercooled liquids \cite{larson,onuki,ferry,onuki2}, and shear thinning
curves are known to be well represented   by expressions
similar to equation (\ref{fit}). 
It is then tempting to interpret our results in terms of
shear rate and stress, using the analogy developed 
at the end of section \ref{pspin}. It is easily seen that
for high ``shear rates'' (defined as 
$\dot{\gamma} \equiv \epsilon/t_\alpha$), equation (\ref{fit}) with $\beta=2$
 implies the relation $t_\alpha \sim \dot{\gamma}^{-2/3}$. 
Remarkably, this is 
very similar to the shear thinning relationship 
observed in polymeric systems 
(for other systems a variety of similar relationships 
can be observed, with shear
thinning exponents that are usually between $-1/2$ and $-1$). 

Note finally  that at very strong drive, the 
relaxation time becomes of the order
of the microscopic time ($t_B \sim 1$ in reduced units and depends only 
slightly on temperature),
the system being Newtonian again.

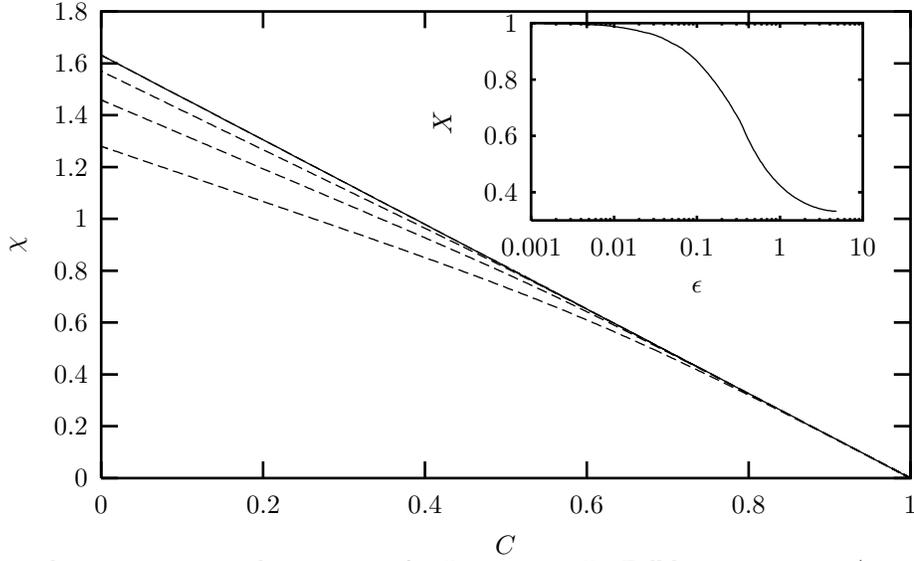
\begin{figure}[t]
\begin{center}
\vspace*{0.5cm}
\input{fdt.613.tex}
\caption{Integrated response vs. correlation curves for $T=0.613>T_c$.
Full line: asymptotic ($\epsilon = 0$) analytical curve.
Dashed lines (from bottom to top) $\epsilon=0.333$, 0.143, 0.05, 0.
Inset: behavior of the FDR as a function of the drive $\epsilon$
for $T=0.613$.}
\label{inset} 
\end{center}
\end{figure}

We now turn to the study of the fluctuation dissipation relation
in the driven system. In the absence of drive, the FDT holds
for $T>T_c$. Hence
a  plot of the integrated response $\chi(\tau)\equiv
\int_0^{\tau} R(\tau') d\tau'$ as a function of $C$ gives a straight line with
slope  $-1/T$.
In figure \ref{inset} we show such a  plot for several strengths of
the drive $\epsilon$. 
We immediately note that the $\chi$ vs. $C$  curve is, for
non-zero drive, well approximated by 
a broken line, with a first piece having a slope  $-1/T$
(corresponding to the fast relaxation) and a second piece 
that displays a smaller slope, denoted in the following by
 $-1/T_{\text{eff}}$ by analogy  
with the analytical results of the above section.

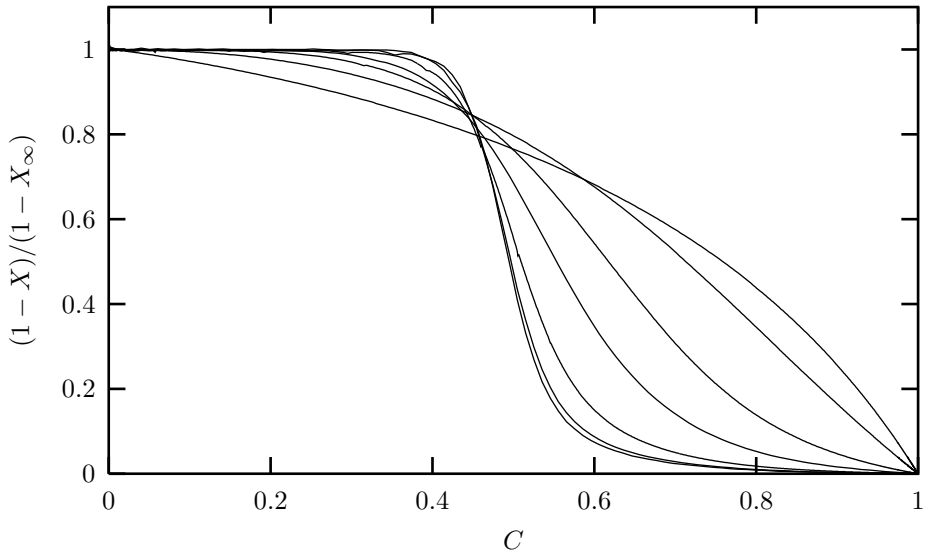
\begin{figure}
\begin{center}
\input{x.62.tex}
\vspace*{0.5cm}
\caption{Normalized fluctuation-dissipation ratio $X$ versus the
  correlation $C$ 
for $\epsilon = 0.00141$, 0.0141, 0.0448, 0.143, 0.333, 1, 5 (from bottom
to top at $C=0.8$). The temperature is $T=0.62$.}
\label{x.62}
\end{center}
\end{figure} 

To make the assumption of a ``two-temperature regime'' very clear,
we plot in Figure \ref{x.62} the normalized  
fluctuation-dissipation ration (FDR)
\begin{equation}
X \equiv \frac{TR(\tau)}{dC(\tau)/d\tau}
\end{equation}
(the normalized derivative of the curves \ref{inset}) 
for several values of the drive. 
The numerical solution shows clearly that, although the
effective temperature 
$T_{\text{eff}}$ is  unambiguously defined
only in the asymptotic limit ($T\rightarrow T_c,  \epsilon\rightarrow 0$),
the ``two straight lines'' approximation is very good even 
at finite driving forces and of course improves as $\epsilon\rightarrow 0$.

The violation of the FDT in a driven system is an important
concept, since it quantifies the deviation from 
the Boltzmann weight in the phase space distribution of the system.
The inset of  figure \ref{inset} shows how the ratio
$X = T_{\text{eff}}/T$ that 
characterizes this violation changes with the driving force.
For small drives, the deviation from one is very small, which 
allows  to define a zone in the plane $(\epsilon,T)$ where
the ``linear response'' (in the sense of irreversible thermodynamic
theory) holds.
This zone happens to coincide with the ``linear regime''
defined from the rheological point of view.
This suggests that there is a direct link between 
a non-linear response in rheological
measurements and a non-equilibrium behavior
from the statistical mechanics point of view.
Note also that one still might study the linear rheology (typically
the behavior of $G^\ast(\omega)$)
of an aging (and hence out of equilibrium) 
system, see \cite{suzanne}.

\subsection{Below the glass transition temperature, $T<T_c$}

Below $T_c$, an undriven 
 system never equilibrates, the
$\alpha$-relaxation time grows with the waiting time elapsed after the 
quench into the glassy phase, and the system never becomes stationary.
If we now quench the system with driving forces acting on it,
it turns out that the $\alpha$-relaxation time becomes finite,
and the system eventually becomes stationary \cite{jorge,CS,horner}.
For vanishingly small drive, however, $t_\alpha$ will again diverge.
The dotted lines in figure \ref{div}-a show this divergence of
$t_\alpha$ with decreasing drive.
For temperatures just below $T_c$, the times at first diverge 
like $\epsilon^{-2}$: the same regime was already noted
above $T_c$.
Still decreasing $\epsilon$, they then cross over to another, 
faster divergence.
Numerically, we find that $t_\alpha (\epsilon \ll 1, T< T_c) \sim
\epsilon^{-\alpha(T)}$, with $\alpha(T)$ slowly (if at all) dependent on
 temperature.
For instance, at $T=0.3$, we obtain $\alpha \simeq 3.19$.
To summarize, below $T_c$, the system does not have a ``Newtonian''
or linear response region, and 
exhibits shear thinning $t_\alpha \sim \epsilon^{-\alpha}$.
This power law divergence is also found numerically
and in a rather involved analytical treatment
 by Horner in the case
of the forced particle in a random potential, where 
 it is called a ``creep'' behavior \cite{horner}.

\begin{figure}
\begin{center}
\input{fdt.45.tex}
\vspace*{0.5cm}
\caption{Integrated response vs. correlation curves for $T=0.45<T_c$.
Full line: asymptotic ($\epsilon = 0$) analytical curve.
Dashed lines (from bottom to top) $\epsilon=0.333$, 0.143, 0.0442.}
\label{fdt.45}
\end{center}
\end{figure}
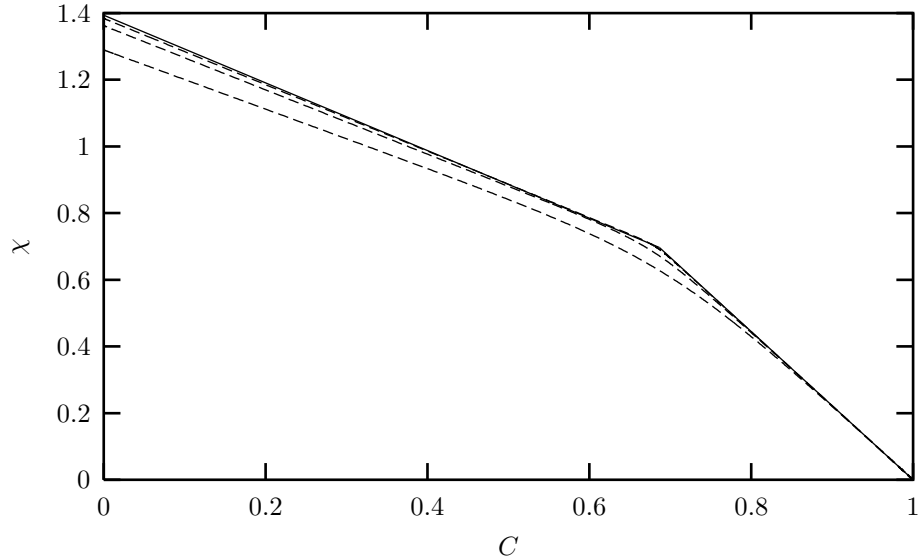 

Turning now to FDT, the situation is quite different from the one
at $T>T_c$ (figure
\ref{fdt.45}). The ``two-temperature scenario'' 
persists even for vanishingly small
drives, giving a clear demonstration of the difference between  
a  glass driven to stationarity 
and an equilibrium system. 
It is very reminiscent of what happens in the aging (undriven)
situation: we already noted that the limiting effective
temperatures were the same in both situations.

\section{Discussion: the temperature-drive ``phase diagram'' }

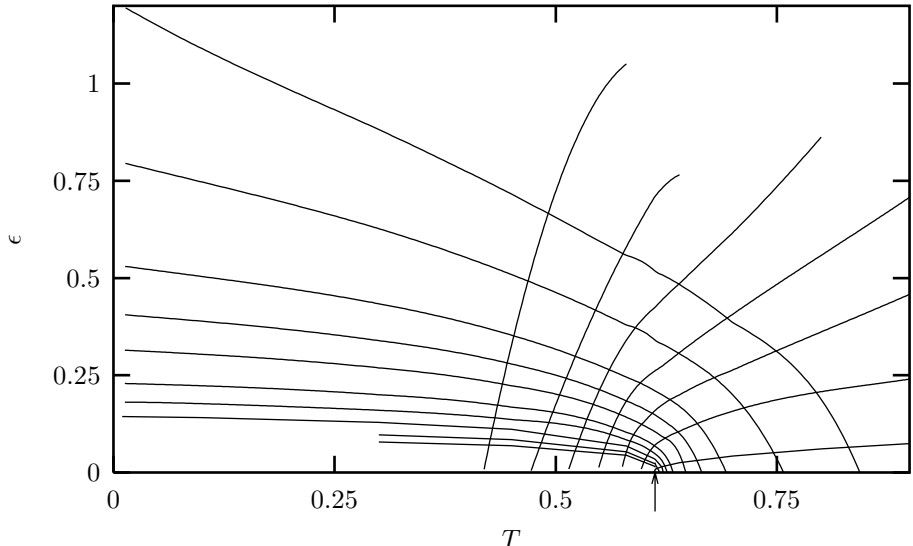
\begin{figure}
\begin{center}
\input{iso.tex} 
\caption{2D view of the glass transition. Curves bent to the left are
the iso-$t_\alpha$, curves bent to the right are the iso-$X$ (see text).
The critical temperature is indicated by the arrow. Times are
$t_\alpha =5$, 10, 25, 50, ..., 5000 (from top to bottom), and
$X=0.4$, 0.5, 0.6, 0.7, 0.8, 0.9, 0.99 (from left to right).}
\label{iso}
\end{center}
\end{figure}

A convenient synthetic representation of the main results obtained 
in this paper is given in figure \ref{iso}, which represents a ``phase
diagram'' of the system in the plane $(\epsilon,T)$. In such a
two dimensional representation, the usual glass phase is restricted
to the segment $T<T_c$ of the horizontal axis. Other points in the
plane corresponds to states which respect time translation invariance, 
and can be characterized by their relaxation time $t_\alpha$ and
their fluctuation dissipation ratio. The two families of curves 
drawn in figure \ref{iso} correspond indeed to constant values 
of $t_\alpha$ (``iso-$t_\alpha$'') and constant values of  the 
long-time FDR
(``iso-$X$''). From these curves, it is manifest that when the 
drive is taken into account, the influence of the glassy phase extends far
beyond the horizontal axis, in the sense that systems with 
values of $X$ characteristic of the glass are found at high temperatures
for finite drives.
The effect of shear is in a sense paradoxical, because at the same
time it makes a glass ``liquid-like'' (by fluidifying it), and 
it brings about typical glass features  (two temperatures)
to a supercooled liquid which otherwise would be at equilibrium. 

Our two dimensional phase diagram is obviously
  closely related to  the  (Load, Temperature) phase diagram 
proposed by   Liu and Nagel in an attempt to rationalize the similarities 
in behavior between  glasses and 
granular matter \cite{nagel}.
In practice, the glass transition temperature
is defined as the temperature where the viscosity 
(or relaxation time) exceeds a 
predefined threshold. Hence the surface drawn by Liu and Nagel to delimit 
``jammed'' (or glassy) 
and ``unjammed'' (or liquid) states would correspond  in our graph
to an iso-$t_\alpha$ line.

The ``near equilibrium'' region of our phase diagram can be defined 
for example as  the zone where $X>0.99$ (note the 
initially flat behavior
of $X(\epsilon)$ in the inset of figure \ref{inset}).
In this region,
 the usual near-equilibrium concepts are applicable. As the glass
transition is approached, this region becomes  
restricted to weaker and weaker  driving forces, and the
two-temperature behavior typical 
of the glassy phase is observed. In the further limit of  
 very large  (probably unphysical in many cases) driving strengths,
 when the relaxation time 
scale becomes of the order of the microscopic time,
 the concept of $T_{\text{eff}}$ becomes of course 
badly defined.

\begin{figure}
\begin{center}
\hspace*{1.cm}
\psfig{figure=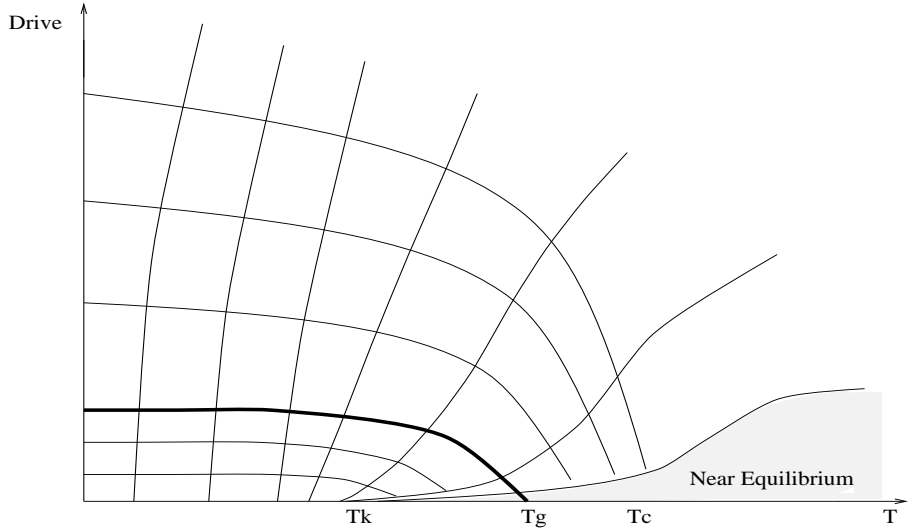,height=7cm,width=12cm} 
\caption{Same as figure \ref{iso}, 
but with the activated processes taken into account (schematic).}
\label{F2}
\end{center}
\end{figure}

As mentioned above, the prediction  of a sharp, purely dynamic
glass transition is an unrealistic feature of any theory not taking
into account activated processes. 
It seems healthy at this point to discuss in some detail what
kind of new features are to be expected in real situations,
when the effect of activated processes is present. 
In figure \ref{F2} we sketch the  modifications that can be  expected
in a realistic system in which the dynamic transition is
smeared by activated processes.
To begin with, at zero drive there is a finite  equilibration time
at all temperatures above  $T_k$ ($T_k<T_c$), the (so-defined)  Kauzmann
temperature, which may or may not be zero. 
Given a system above $T_k$ there will, in principle,
 always be a sufficiently small
driving force  such that FDT still holds.
However, not all of the  drive-temperature plane is accessible in
stationary (non-aging) conditions. Given an experimental situation, there is a
level
 (the thick line in Fig. \ref{F2}) 
below which the system does not have the time to  become
stationary. this is the ``glass transition'' (in fact, a crossover)
 as it happens in practice,
and the zero-drive intercept of  this line is usually denoted the
``glass transition temperature''  $T_g$.

From what we have learned from the idealized case, we might expect
that  above $T_g$, in conditions such that the  $\alpha$-relaxation
time and therefore
the viscosity
are already large, there will be a well established
two-temperature
regime as soon as a moderately weak driving is turned on.
The consequence is that 
in the supercooled liquid regime for instance, a small shear rate 
will be sufficient to give at the same time a non-linear
rheological behavior (as seen in the 
MD simulations of reference \cite{onuki}),
and {\it an effective temperature different from that
of the thermostat}.
Such an effect would also be accessible in an experiment.
We stress that an effective temperature could be 
measured in that case {\it in a stationary state}
making the measurements far easier than in an
aging experiment \cite{israeloff}.

Below the ``glass transition'' line, on the contrary, we expect
that, as discussed in section \ref{yield},  history dependent
 effects will become predominant, with effects like yield stress
 or hysteresis that
cannot be accounted for within mean-field like theoretical schemes.
 
Having pointed out all these differences, the basic suggestion 
of a two steps, two temperatures relaxation behavior remains,
and is open for numerical and experimental checks.

\section{Conclusions and perspectives}

In this paper, we have studied, within the framework
of  the mean-field (or mode-coupling) approximation,
 the influence of an external drive
on a system undergoing a glass transition. 
The main results of the approach should, we believe,
not depend  too strongly on  
 the method of approximation. Hence we have constantly
tried to interpret the results within the context of a rheological
experiment on a system with anomalous rheology, although such an interpretation
remains of course rather tentative.

In this language, we can summarize our results as follows. 
 If a fluid having, in equilibrium,   
two well-separated relaxation time scales is gently driven (e.g. sheared)
the slow structural relaxations are accelerated (shear thinning).
  This acceleration is accompanied by 
the appearance of an effective temperature for the 
 slow structural degrees of freedom, while the fast degrees of
 freedom
(phonons, etc) are still at the bath temperature.  
Each  temperature is  associated with fluctuations at each
 time scale, and is
well-defined if the drive is gentle enough that the time scales still remain
separated \cite{sollich0}.
When the two temperatures differ very little, ``near equilibrium'' assumptions
becomes justified: this regime was shown to coincide with 
the usual ``linear regime'' of steady-state rheology. 
For stronger drive, a power law decay of the relaxation
time with increasing drive is observed, as is the case
in many complex fluids. Interestingly, the associated 
``shear thinning'' exponent is found to be $-2/3$, and seems
to be quite robust with respect to variations in the model.
This could be an indication of why so many shear thinning exponents 
in real systems are found in the range [-1/2,-1].

The obvious weakness of the approach, which is intrinsic
to the perturbative scheme adopted, is that 
the effect of activated processes 
can only be described at a qualitative level. Hence
a number of interesting phenomena observed at weak driving forces (e.g.
yield stress, dilatancy, hysteresis) cannot be addressed 
analytically.

In principle, the present study could
be extended to equations with many coupled
spatial modes.   One can even go further and consider richer resummation
schemes, like for example the  self consistent screening approximation
\cite{bouchaud}. 
Moreover, other resummations  like applying to the driven case a
dynamical version of the hypernetted-chain equation 
(in the spirit of the work on
glasses of M\'ezard and Parisi \cite{Mepa}) can be envisaged.
In this way, one could study a true shear applied to the largest spatial
mode, and calculate how the energy cascades  to the shorter
wavelengths.
The two-temperature ansatz can be shown to close \cite{statphys}
independently of the resummation scheme, with the important property
that there are still in the small drive limit 
only two temperatures shared by all spatial modes.
The implementation of these improvements may then allow to extract
actual numbers starting from a realistic microscopic theory. 
However,
it should be born in mind the non-perturbative activated processes
would still remain unaccessible to these more sophisticated 
computational schemes.

\section*{Acknowledgments}
The numerical calculations presented in this work were carried
out at the PSMN of ENS-LYON.
We thank Arnulf Latz for providing us with a copy of the program described
in Reference \cite{arnulf}, and F. Thalmann for discussions.
J. K. was partially supported by the ``Programme Th\'ematique Mat\'eriaux,
  R\'egion Rh\^one-Alpes''.

\section*{Appendix: calculation of the power input}
The quantity to estimate is 
\begin{equation}
P \equiv \overline{\langle \frac{1}{N} \sum_{i=1}^N f_i^{\text{drive}}
\dot{s_i} \rangle }.
\end{equation}
To compute this quantity, we introduce a 
current 
$h$, and a generating functional  $Z$ such that
\begin{equation}
Z \equiv \overline{\langle \exp \left( \int dt \sum_i h_i(t)
f_i^{\text{drive}} \dot{s_i} \right) \rangle }.
\end{equation}
Then by definition
\begin{equation}
P(t) = \frac{1}{N} \sum_{i=1}^N \left[  \frac{\delta Z}{\delta h_i(t)} 
\right] _{h=0} .
\end{equation}
 $Z$ can be obtained from the usual 
Martin-Siggia-Rose (MSR) formalism,
and the average over the random variables $\tilde{J}$ gives the extra-term 
\begin{equation}
\begin{aligned}
\frac{k \epsilon^2}{4N^{k-1} } \sum_{i,j_1,\cdots,j_{k-1}}\int dt dt' 
 & [ s_{j_1}(t) 
s_{j_1}(t') \cdots s_{j_{k-1}} (t)  s_{j_{k-1}} (t') ] \times
\\
& [ i\hat{s_i}(t)  i\hat{s_i}(t') + h_i(t)\dot{s_i}(t)
h_i(t')\dot{s_i}(t')+2 i\hat{s_i}(t') h_i(t)\dot{s_i}(t)]
\end{aligned}
\end{equation}
in the effective MSR Lagrangian.
Finally, we obtain for $P$:
\begin{equation}
P(t) = \int {\cal D}s {\cal D} \hat{s} \exp(L) 
\frac{k \epsilon^2}{4N^{k} } \sum_{i,j_1,\cdots,j_{k-1}}\int dt' 
 [ s_{j_1}(t) 
s_{j_1}(t') \cdots s_{j_{k-1}} (t)  s_{j_{k-1}} (t') ] 
[ 2 i\hat{s_i}(t') \dot{s_i}(t)],
\end{equation}
where $L$ is the usual MSR Lagrangian for the $p$-spin model.
The integral is estimated at its saddle point, yielding,
\begin{equation}
P(t)= \frac{k \epsilon^2}{2} \int_{-\infty}^t dt' C(t,t')^{k-1} 
\frac{\partial R(t,t')}{\partial t}.
\end{equation}
which reduces for time translation invariant systems to 
\begin{equation}
P= \frac{k \epsilon^2}{2} \int_0^{+\infty} d\tau C(\tau)^{k-1} 
\frac{d R(\tau)}{d \tau}.
\end{equation}

\end{document}

%% file: corrT=.613.tex
\begingroup%
  \makeatletter%
  \newcommand{\GNUPLOTspecial}{%
    \@sanitize\catcode`\%=14\relax\special}%
  \setlength{\unitlength}{0.1bp}%
\begin{picture}(3600,2160)(0,0)%
\special{psfile=corrT=.613 llx=0 lly=0 urx=720 ury=504 rwi=7200}
\put(1925,50){\makebox(0,0){$\tau$}}%
\put(100,1180){%
\special{ps: gsave currentpoint currentpoint translate
270 rotate neg exch neg exch translate}%
\makebox(0,0)[b]{\shortstack{$C(\tau)$}}%
\special{ps: currentpoint grestore moveto}%
}%
\put(3196,200){\makebox(0,0){$10^6$}}%
\put(2688,200){\makebox(0,0){$10^4$}}%
\put(2179,200){\makebox(0,0){$10^2$}}%
\put(1671,200){\makebox(0,0){$10^0$}}%
\put(1163,200){\makebox(0,0){$10^{-2}$}}%
\put(654,200){\makebox(0,0){$10^{-4}$}}%
\put(350,2060){\makebox(0,0)[r]{1}}%
\put(350,1708){\makebox(0,0)[r]{0.8}}%
\put(350,1356){\makebox(0,0)[r]{0.6}}%
\put(350,1004){\makebox(0,0)[r]{0.4}}%
\put(350,652){\makebox(0,0)[r]{0.2}}%
\put(350,300){\makebox(0,0)[r]{0}}%
\end{picture}%
\endgroup
 

%% file: scaleaboveTc.tex
\begingroup%
  \makeatletter%
  \newcommand{\GNUPLOTspecial}{%
    \@sanitize\catcode`\%=14\relax\special}%
  \setlength{\unitlength}{0.1bp}%
\begin{picture}(3600,2160)(0,0)%
\special{psfile=scaleaboveTc llx=0 lly=0 urx=720 ury=504 rwi=7200}
\put(2000,50){\makebox(0,0){$\epsilon / \epsilon_0$}}%
\put(100,1180){%
\special{ps: gsave currentpoint currentpoint translate
270 rotate neg exch neg exch translate}%
\makebox(0,0)[b]{\shortstack{$(t_\alpha - t_B) / (t_0 - t_B)$}}%
\special{ps: currentpoint grestore moveto}%
}%
\put(3450,200){\makebox(0,0){$10^3$}}%
\put(2967,200){\makebox(0,0){$10^2$}}%
\put(2483,200){\makebox(0,0){$10$}}%
\put(2000,200){\makebox(0,0){$10^0$}}%
\put(1517,200){\makebox(0,0){$10^{-1}$}}%
\put(1033,200){\makebox(0,0){$10^{-2}$}}%
\put(550,200){\makebox(0,0){$10^{-3}$}}%
\put(500,1960){\makebox(0,0)[r]{$10^0$}}%
\put(500,1628){\makebox(0,0)[r]{$10^{-1}$}}%
\put(500,1296){\makebox(0,0)[r]{$10^{-2}$}}%
\put(500,964){\makebox(0,0)[r]{$10^{-3}$}}%
\put(500,632){\makebox(0,0)[r]{$10^{-4}$}}%
\put(500,300){\makebox(0,0)[r]{$10^{-5}$}}%
\end{picture}%
\endgroup
 

%% file: fdt.613.tex
\begingroup%
  \makeatletter%
  \newcommand{\GNUPLOTspecial}{%
    \@sanitize\catcode`\%=14\relax\special}%
  \setlength{\unitlength}{0.1bp}%
\begin{picture}(3600,2160)(0,0)%
\special{psfile=fdt.613 llx=0 lly=0 urx=720 ury=504 rwi=7200}
\put(2645,1022){\makebox(0,0){$\epsilon$}}%
\put(1720,1644){%
\special{ps: gsave currentpoint currentpoint translate
270 rotate neg exch neg exch translate}%
\makebox(0,0)[b]{\shortstack{$X$}}%
\special{ps: currentpoint grestore moveto}%
}%
\put(3270,1172){\makebox(0,0){10}}%
\put(2958,1172){\makebox(0,0){1}}%
\put(2645,1172){\makebox(0,0){0.1}}%
\put(2333,1172){\makebox(0,0){0.01}}%
\put(2020,1172){\makebox(0,0){0.001}}%
\put(1970,2016){\makebox(0,0)[r]{1}}%
\put(1970,1803){\makebox(0,0)[r]{0.8}}%
\put(1970,1591){\makebox(0,0)[r]{0.6}}%
\put(1970,1378){\makebox(0,0)[r]{0.4}}%
\put(1925,50){\makebox(0,0){$C$}}%
\put(100,1180){%
\special{ps: gsave currentpoint currentpoint translate
270 rotate neg exch neg exch translate}%
\makebox(0,0)[b]{\shortstack{$\chi$}}%
\special{ps: currentpoint grestore moveto}%
}%
\put(3450,200){\makebox(0,0){1}}%
\put(2840,200){\makebox(0,0){0.8}}%
\put(2230,200){\makebox(0,0){0.6}}%
\put(1620,200){\makebox(0,0){0.4}}%
\put(1010,200){\makebox(0,0){0.2}}%
\put(400,200){\makebox(0,0){0}}%
\put(350,2060){\makebox(0,0)[r]{1.8}}%
\put(350,1864){\makebox(0,0)[r]{1.6}}%
\put(350,1669){\makebox(0,0)[r]{1.4}}%
\put(350,1473){\makebox(0,0)[r]{1.2}}%
\put(350,1278){\makebox(0,0)[r]{1}}%
\put(350,1082){\makebox(0,0)[r]{0.8}}%
\put(350,887){\makebox(0,0)[r]{0.6}}%
\put(350,691){\makebox(0,0)[r]{0.4}}%
\put(350,496){\makebox(0,0)[r]{0.2}}%
\put(350,300){\makebox(0,0)[r]{0}}%
\end{picture}%
\endgroup
 

%% file: x.62.tex
\begingroup%
  \makeatletter%
  \newcommand{\GNUPLOTspecial}{%
    \@sanitize\catcode`\%=14\relax\special}%
  \setlength{\unitlength}{0.1bp}%
\begin{picture}(3600,2160)(0,0)%
\special{psfile=x.62 llx=0 lly=0 urx=720 ury=504 rwi=7200}
\put(1925,50){\makebox(0,0){$C$}}%
\put(100,1180){%
\special{ps: gsave currentpoint currentpoint translate
270 rotate neg exch neg exch translate}%
\makebox(0,0)[b]{\shortstack{$(1-X)/(1-X_\infty)$}}%
\special{ps: currentpoint grestore moveto}%
}%
\put(3450,200){\makebox(0,0){1}}%
\put(2840,200){\makebox(0,0){0.8}}%
\put(2230,200){\makebox(0,0){0.6}}%
\put(1620,200){\makebox(0,0){0.4}}%
\put(1010,200){\makebox(0,0){0.2}}%
\put(400,200){\makebox(0,0){0}}%
\put(350,1900){\makebox(0,0)[r]{1}}%
\put(350,1580){\makebox(0,0)[r]{0.8}}%
\put(350,1260){\makebox(0,0)[r]{0.6}}%
\put(350,940){\makebox(0,0)[r]{0.4}}%
\put(350,620){\makebox(0,0)[r]{0.2}}%
\put(350,300){\makebox(0,0)[r]{0}}%
\end{picture}%
\endgroup
 

%% file: fdt.45.tex
\begingroup%
  \makeatletter%
  \newcommand{\GNUPLOTspecial}{%
    \@sanitize\catcode`\%=14\relax\special}%
  \setlength{\unitlength}{0.1bp}%
\begin{picture}(3600,2160)(0,0)%
\special{psfile=fdt.45 llx=0 lly=0 urx=720 ury=504 rwi=7200}
\put(1925,50){\makebox(0,0){$C$}}%
\put(100,1180){%
\special{ps: gsave currentpoint currentpoint translate
270 rotate neg exch neg exch translate}%
\makebox(0,0)[b]{\shortstack{$\chi$}}%
\special{ps: currentpoint grestore moveto}%
}%
\put(3450,200){\makebox(0,0){1}}%
\put(2840,200){\makebox(0,0){0.8}}%
\put(2230,200){\makebox(0,0){0.6}}%
\put(1620,200){\makebox(0,0){0.4}}%
\put(1010,200){\makebox(0,0){0.2}}%
\put(400,200){\makebox(0,0){0}}%
\put(350,2060){\makebox(0,0)[r]{1.4}}%
\put(350,1809){\makebox(0,0)[r]{1.2}}%
\put(350,1557){\makebox(0,0)[r]{1}}%
\put(350,1306){\makebox(0,0)[r]{0.8}}%
\put(350,1054){\makebox(0,0)[r]{0.6}}%
\put(350,803){\makebox(0,0)[r]{0.4}}%
\put(350,551){\makebox(0,0)[r]{0.2}}%
\put(350,300){\makebox(0,0)[r]{0}}%
\end{picture}%
\endgroup
 

%% file: iso.tex
\begingroup%
  \makeatletter%
  \newcommand{\GNUPLOTspecial}{%
    \@sanitize\catcode`\%=14\relax\special}%
  \setlength{\unitlength}{0.1bp}%
\begin{picture}(3600,2160)(0,0)%
\special{psfile=iso llx=0 lly=0 urx=720 ury=504 rwi=7200}
\put(1950,50){\makebox(0,0){$T$}}%
\put(100,1180){%
\special{ps: gsave currentpoint currentpoint translate
270 rotate neg exch neg exch translate}%
\makebox(0,0)[b]{\shortstack{$\epsilon$}}%
\special{ps: currentpoint grestore moveto}%
}%
\put(2950,200){\makebox(0,0){0.75}}%
\put(2117,200){\makebox(0,0){0.5}}%
\put(1283,200){\makebox(0,0){0.25}}%
\put(450,200){\makebox(0,0){0}}%
\put(400,1767){\makebox(0,0)[r]{1}}%
\put(400,1400){\makebox(0,0)[r]{0.75}}%
\put(400,1033){\makebox(0,0)[r]{0.5}}%
\put(400,667){\makebox(0,0)[r]{0.25}}%
\put(400,300){\makebox(0,0)[r]{0}}%
\end{picture}%
\endgroup
 

%% file: revised.bbl
\begin{references}
\bibitem{sollich0} The hopping temperature that plays an important
 role of 
in the SGR model\cite{sollich} is more like the effective temperature
 of the slow modes here. It is not unlikely that a precise connection 
will be possible.
\end{references}
